\documentstyle[12pt]{article}
\begin{document}
\setlength{\unitlength}{1mm}

{\hfill   January 1997 }

{\hfill    WATPHYS-TH-97/01}

{\hfill    hep-th/9701106} \vspace*{2cm} \\
\begin{center}
{\Large\bf Entropy of Schwarzschild black hole and }
\end{center}
\begin{center}
{\Large\bf
string-black hole 
correspondence } 
\end{center}
\begin{center}
Sergey N.~Solodukhin\footnote{e-mail: sergey@avatar.uwaterloo.ca}
\end{center}
\begin{center}
{\it Department of Physics, University of Waterloo, Waterloo, Ontario N2L 3G1, 
Canada}  
\end{center}
\vspace*{1cm}
\begin{abstract}
The {\it string-black hole} correspondence is considered in the context
of the correspondence principle proposed recently by Horowitz and Polchinski
\cite{4}. We demonstrate that the entropy of string states and the
entropy of a Schwarzschild black hole can be matched including
the subleading terms which depend on mass logarithmically.
We argue the necessity to include the string interaction (with coupling $g$)
in the consideration and propose the $g^2$-dependent modification
of the string entropy. The matching of it with the entropy of Schwarzschild
black hole is analyzed.  We also discuss a possible scenario 
when the entropy of a Schwarzschild black hole  appears entirely 
as effect of the interaction.
\end{abstract}
\begin{center}
{\it PACS number(s): 04.70.Dy, 04.62.+v}
\end{center}
\vskip 1cm
\newpage
\baselineskip=.8cm
An intriguing idea proposed by Susskind \cite{1}
is to identify the states of a black hole with the highly
excited states of a fundamental string. A heuristic argument in support
of this suggestion
is that the number of states for both the hole and the
string grows rapidly  as a function of mass; and as the
string coupling increases, the size of a string state becomes smaller than
its Schwarzschild radius, hence any such  state must collapse to form a
black hole. Recently, remarkable progress has been made towards
establishing this correspondence for an extreme black hole:
the relevant string states are BPS states characterized by charges 
identical to those of the extreme black hole \cite{2}. Their 
entropy  is then identical to the Bekenstein-Hawking
entropy defined by the horizon area \cite{2} (see review in \cite{3}).

The attempt to extend 
this correspondence to non-extreme black holes
\cite{1}, \cite{2''}, \cite{2'}, \cite{3'}
runs into an obvious problem: the number of states of the black hole grows
as $e^{M^2}$ while that of the string  as $e^M$. To resolve this discrepancy, 
Horowitz and Polchinski \cite{4} have  recently proposed a
 correspondence principle: the entropy of string states and the
 black hole entropy should be  matched only for a single
(critical) value of the string coupling $g=g_{cr}$, for which the string
 size becomes of the order of the Schwarzschild radius.
Therefore, varying the coupling $g$ from zero to $g_{cr}$ we travel from the 
{\it free-string phase} to the phase of the collapsing strings which ends
by the formation of the hole.

In this note we make a few comments regarding the {\it string-black hole}
correspondence for Schwarzschild black hole 
and, in particular, arrive at a conclusion that black hole
states can be naturally identified with the states of {\it perturbatively 
interacting string} rather than with the  {\it free} states.

\bigskip

We start with considering a free string. The 
number of its states at the excitation 
level $N$ is given by  \cite{4'}
\begin{equation}
d(N)=e^{2\pi\sqrt{{c\over 6}N}}N^{-B}~~,
\label{1}
\end{equation}
where $c$ is the effective two-dimensional central charge, 
while mass of the excitation is
\begin{equation}
 M^2_s={N\over l^2_s}~~,
\label{2}
\end{equation}
where $l_s$ is the string size, $l_s\sim\sqrt{\alpha '}$.  The exponent
$B$ in (\ref{1}) is universally related to $D_\bot$, the effective number of space-time 
dimensions for transverse string oscillations (or, equivalently, the number 
of uncompactified bosonic degrees of freedom in the worldsheet CFT 
\cite{Vafa}): $B={1\over 4}(3+D_\bot )$. We have \cite{4'} $B={27\over 4}$
for bosonic string and $B={11\over 4}$ for type-II superstring and
heterotic string.

Define a macroscopic state of the string system by fixing the mass $M_s$
(or, equivalently, $N$ in accord with (\ref{2})). Then it is realized
by $d(N)$ microstates of the string with the entropy
\begin{eqnarray}
&&S_s=\ln d(N)=2\pi\sqrt{c\over 6}\sqrt{N}-B \ln N \nonumber \\
&&=2\pi\sqrt{c\over 6}(l_sM_s)-2B\ln (l_sM_s)~~,
\label{3}
\end{eqnarray}
where in the last line we have used (\ref{2}). We see that the subleading
term in (\ref{1}) gives rise to a subleading term  in the string 
entropy (\ref{3}) which  depends on mass logarithmically. 
A natural question arises  if there exists an analogous term in the entropy
in the black hole phase?

A black hole in  string theory arises 
as a solution of the low-energy action which can be isolated as
the lowest order term in the expansion  
\begin{equation}
W_{eff}=-{1\over16\pi G} \left( \int_{M^4} R +\int_{M^4} 
{\cal L} (\phi , A_\mu , g_{\mu\nu}, ...)
\right) -\sum_{k=0}^{\infty} (gl_s)^{2k} W_{k}~~,
\label{4}
\end{equation}
where $G=g^2l^2_s$ is the string induced gravitational constant,
the Planck distance $l_{pl}$ is defined as $G=l_{pl}^2$.
${\cal L} (\phi , A_\mu , g_{\mu\nu}, ...)$ is the Lagrangian for the (super)
multiplet of matter fields appearing in the low-energy approximation. 
Other terms ($\sim (gl_s)^{2k}$) 
in the expression (\ref{4}) are presumably rather complicated
non-local functionals. Their influence on a low-energy solution
results in some ``quantum deformation'' of the solution. In other words,
they are responsible for the quantum back-reaction effects.

We will be considering here only Schwarzschild black holes. Generally, they 
are  non-extreme black hole geometries characterized
by a single  dimensional parameter - the Schwarzschild radius $r_+$.
It is related to the black hole mass $M_{bh}$ as $r_+=2M_{bh} l^2_{pl}$.
Each term in (\ref{4}) which depends on 
curvature   contributes to the entropy of  
the hole. The Einstein-like term  gives rise to the
standard Bekenstein-Hawking expression relating the entropy with the horizon
area. In general, the contribution of other terms ($\sim (gl_s)^{2k}$)
in (\ref{4}) to the entropy  is not easily recovered.
However, for a Schwarzschild hole this can be accomplished as follows.
 The  Schwarzschild geometry is characterized by a single
 dimensional parameter 
$r_+$, and hence   one can apply  scaling and dimensionality
arguments.  They give rise to the entropy
of the Schwarzschild black hole which in the leading (with respect to $r_+$)
order has the form:
\begin{equation}
S_{bh}=\pi{r^2_+\over l^2_{pl}}-c_1\ln ({r_+\over {\bf \mu}} )~~,
\label{5}
\end{equation}
where $c_1$ is the four-dimensional central charge which comes from 
the  integrated $4D$ conformal anomaly for the zero-mass fields in the
theory (\ref{4}):
$\int d^4x\sqrt{g} T^\mu_\mu=c_1$; $\mu^{-1}$ is a mass scale.
For a massless theory with $N_0$ scalars, $N_{1/2}$ Majorana fermions,
$N_1$ vectors, $N_{3/2}$ spin-$3/2$ fermions, $N_2$ gravitons, and
$N_A$ rank-two antisymmetric tensor fields, the coefficient $c_1$ reads 
\cite{*}, \cite{**}
\begin{equation}
c_1={\chi\over 90}(-N_0-{7\over 4}N_{1/2}+13 N_1+{233\over 4}N_{3/2}-212 N_2-
91 N_A )~~,
\end{equation}
where $\chi$ is the Euler number, for the Schwarzschild black hole it is equal
to $2$.

 In the microcanonical ensemble the black hole entropy can be calculated 
as minus the action functional considered on the Euclidean black hole 
instanton. Thus, the classical Bekenstein-Hawking entropy in (\ref{5})
arises from the first (low-energy) term in the effective action (\ref{4}) while
the series $\sum_{k=0}^\infty (gl_s)^{2k}W_k$ results in some corrections.
The non-local part of the 
first term in the series, $W_0$, can be interpreted as due to the one-loop
quantization of the low-energy theory. Its contribution to the entropy is
isolated by the scaling arguments\footnote{ The contribution to the entropy of 
other terms in the series is proportional to $({gl_s\over r_+})^{2k}\simeq
({1\over gl_sM})^{2k},~k=1,2,...$ and is 
omitted in (\ref{5}).}.
 Indeed, considering the quantum part of the effective action (\ref{4})
on the Schwarzschild black hole instanton,
$W_0[g^{sch}_{\mu\nu}(r_+)]$, and performing
the rescaling we find  
$$W_0[g^{sch}_{\mu\nu}(r_+)]=
W_0[\alpha^2g^{sch}_{\mu\nu}({r_+\over \alpha})]=W_0[g^{sch}_{\mu\nu}({r_+\over\alpha})]+
\left( \int d^4x\sqrt{g} T^\mu_\mu \right) \ln \alpha~~
$$
that gives $W_0[g^{sch}_{\mu\nu}(r_+)]=c_1 \ln {r_+\over \mu}+const$ and 
the second term in 
(\ref{5}) (an additive constant is omitted in (\ref{5})).     

It should be noted that the
form (\ref{5}) is quite universal. It  appears in different models
in two \cite{5}, three \cite{6} and four \cite{7} dimensions.
Unfortunately, for a charged black hole the scaling arguments are not
restrictive enough to recover the form of the entropy since there
are more than one dimensional parameter characterizing the geometry
(see, however, \cite{8}).  

A nice thing about (\ref{5}) is that it already
contains the back-reaction effects if the horizon radius $r_+$ is considered as
the radius of the ``quantum-corrected'' black hole. In terms of the
black hole mass
Eq.(\ref{5}) reads\footnote{In general, the relation between the 
horizon radius and
mass $M$ for a ``quantum-corrected'' black hole
might be the following: $r^2_+=4M^2l^4_{pl}+\kappa l^2_{pl}\ln M$
(see last paper in Ref.\cite{5}).
This would modify the coefficient in front of the $\log$ in (\ref{6}):
$c_1\rightarrow c_1-\kappa\pi$.}
\begin{equation}
S_{bh}=4\pi (gl_s M_{bh})^2-c_1\ln (l_s M_{bh})~~,
\label{6}
\end{equation}
where we have used  $l_{pl}=gl_s$ and omitted term 
$\sim \ln( {l_sg^2\over \mu})$.
This is  the entropy in the {\it black hole phase}. The 
comparison of it with the
entropy in the {\it free-string phase} (\ref{3}) shows that both
quantities $S_s$ and $S_{bh}$ have  similar subleading terms while 
their leading behavior is considerably different: $S_s\sim M$, $S_{bh}\sim
M^2$. In order to resolve this, Horowitz and Polchinski \cite{4}
proposed the correspondence principle according 
to which the two expressions (\ref{3}) and (\ref{6}) are matched for a single value
$g=g_{cr}$ of the string coupling constant. This value of the coupling 
corresponds  to the
transition from the string description to the black hole description.
Indeed, as $g$ increases, the string macroscopic state collapses to 
form a  black hole. This is signaled by
the size of the string
becoming of the order of   the Schwarzschild radius, or more precisely,  
$l_s=\sqrt{c\over 6}r_+$.
This identity defines  $g_{cr}$.
Assuming that the mass does not change during the transition, $M_s=M_{bh}$,
we find that the black hole forms at the following value of the string 
coupling 
\begin{equation}
g^{2}_{cr}=\sqrt{c\over 24}(l_sM_{})^{-1}\simeq N^{-1/2}~~.
\label{8}
\end{equation}
For this value of $g$ 
the entropies (\ref{3}) and (\ref{6}) are equal if the $4D$
and $2D$ central charges  are related as $c_1=2B$.
If so, the {\it string-black hole} correspondence may have a wider range of
validity than originally anticipated.

The relation $c_1=2B$ is not automatically satisfied and is a constraint on the low-energy string configuration. As is well known \cite{*}, \cite{**}
the conformal anomaly in four dimensions vanishes 
($c_1=0$) for $N=4,~8$ supergravity 
theories.  The non-vanishing contribution is possible 
from sectors which preserve 
$N=2$ space-time supersymmetry \cite{**}. For a heterotic $N=2$ vacuum the
relevant massless spectrum contains \cite{***}: {\it the gravity multiplet}:
the graviton, two gravitinos and a spin-$1$ Abelian gauge boson (graviphoton)
(its contribution to the conformal anomaly is $c_1=-{11\over 6}$); {\it the 
vector-tensor multiplet}: the dilaton, the rank-two antisymmetric tensor field,
$2$ Majorana fermions, and a vector boson (the total $c_1=-{11\over 6}$);
{\it the vector multiplet}: a gauge boson, $2$ Majorana fermions, and
a complex scalar ($c_1={1\over 6}$); {\it the hypermultiplet}: $4$ scalars and
$2$ Majorana fermions ($c_1=-{1\over 6}$). In type-II theories the dilaton resides
in the {\it tensor multiplet} \cite{88} 
consisting on the antisymmetric tensor, $ 2$ Majorana fermions and $3$ scalars ($c_1=-{13\over 6}$)\footnote{In four dimensions an antisymmetric tensor is 
dual to a scalar field (the axion) and thus the {\it vector-tensor multiplet} 
is dual to an Abelian vector multiplet. On the other hand, the $N=2$
{\it tensor multiplet} is dual to an $N=2$ {\it hypermultiplet}. The conformal
anomaly, however, depends on the field representation and  differs for the
dual multiplets \cite{*}.}. The members of the vector multiplet
are in the adjoint representation of the gauge group $G$. In general we have
$n_v$ vector multiplets and $n_h$ hypermultiplets. The  massless spectrum results in the total 4D central charge
$$
c_1={1\over 6}(n_v-n_h-22)
$$
for heterotic string theory and
$$
c_1={1\over 6}(n_v-n_h-24)
$$
for type-II string theory. In order to compare this result to the coefficient
in front of the subleading term in (\ref{3}) we should note  that
in general the degeneracy of the string states is a product of 
the degeneracies 
coming from  the right-
and left-moving sectors. Therefore, the matching condition for both the type-II
and heterotic strings is $c_1=11$.
  It is satisfied if numbers of the vector- and 
hyper-multiplets are related as
\begin{eqnarray}
&&n_v-n_h=88 ~~~for~~heterotic~~strings \nonumber \\
&&n_v-n_h=90 ~~~for~~type-II~~strings
\label{!}
\end{eqnarray}
The compactification of the $D=10$ heterotic string on a six-dimensional
manifold $T^2 \times K_3$ leads (see for example \cite{***}) 
to the gauge group $G=E_7\times E_8\times U(1)^2$.
This vacuum has $n_v=dim~ G=383$ vector multiplets. Then Eq.(\ref{!})
gives $n_h=255$ for the number of $N=2$ hypermultiplets. 
Such an $N=2$ configuration (if it is actually realizable) gives us an
example when the logarithmic term in the black hole entropy (\ref{5})
has statistical origin as due to the subleading behavior of the 
number of string states (\ref{3}).  However, we should note that
the coefficient $c_1$ in the black hole calculation
seems to depend on the compactification while the coefficient in the
string calculation does not. A possible resolution of this is that the 
matching condition gives us a constraint on the compactification.
On other hand, if the coefficients are not matched we still have 
agreement between two descriptions which now happens for the
string coupling taking the critical value modified by the logarithmic term:
\begin{equation}
g^2_{cr}=\sqrt{c\over 24}{1\over (l_s M)^2}(l_sM+\sigma \ln (l_s M))~~,
\label{g}
\end{equation}
where $\sigma=\sqrt{24\over c}{1\over 2\pi}(c_1-2B)$.

It should be noted that in the above consideration  we were matching
 two quantities 
$S_s$ and $S_{bh}$ which are defined for essentially  different
values of $g$. Indeed, $S_s$ is a free-string quantity ($g=0$), while
$g$ is supposed to be non-zero  in defining $S_{bh}$. 
Extrapolating  $S_s$ to non-zero values of $g$ we get the
 critical 
value $g_{cr}\sim N^{-1/4}$. 
However, there must be an intermediate {\it gravitational
phase} between the {\it free-string phase}
and the {\it black hole phase}, 
when the string states are already attracting but the black hole
is not yet formed. This phase is characterized by the appearance of
the new scale \cite{9} playing the role of an ``order parameter''.
This is, of course, the Planck scale $l_{pl}=gl_s$. In this phase the string 
interaction can   be considered  perturbatively that results in
a $g$-dependent string entropy $S^g_s$. It seems reasonable that
namely  states of the interacting string may form the hole and in 
the critical point (when the string and black hole descriptions coincide)
we must match the black hole entropy $S_{bh}$ and the interacting string 
entropy $S^g_s$ (instead of
the free quantity $S_s$). 
In principle, the string entropy $S^g_s$ can possess some non-perturbative corrections, behaving as $\sim {1\over g^2}$, which are hard to reveal.  
Therefore, in what follows we consider the string entropy $S^g_s$
as a perturbation series with respect to $g$ and discuss its matching
to $S_{bh}$. 

We know that the quantities (\ref{1})-(\ref{3}) are valid 
for free string, $g=0$. When the string interaction is turned on there must
appear $g$-dependent corrections to these formulas. It is reasonable to expect 
that these corrections are controlled by $g^2$. Therefore, by dimensional
arguments \cite{arg} we obtain that such a correction to the entropy
 is determined by the quantity
$(gl_sM_s)^2$. With this correction included,
 the expected expression for the entropy of
the string states for $g\neq 0$ is 
\begin{equation}
S_s^{g}=2\pi\sqrt{c\over 6}(l_sM_s)+4\pi a(gl_sM_s)^2-2B
 \ln \left((l_sM_s)+4\pi a(gl_sM_s)^2
\right)~~,
\label{9}
\end{equation}
where the value of coefficient $a$ can  in 
principle be determined by a precise calculation. 
The perturbatively interacting string  presumably has the same level
structure as the free string. Therefore, the formula (\ref{1}) is still
valid for $g\neq 0$. However, the mass formula (\ref{2}) for the excited states
gets modified as $\sqrt{N}=l_sM_s+a'(gl_sM_s)^2$ ($a=\sqrt{c\over 24}a'$).
These result in the Eq.(\ref{9}).
For 
small $g$ we find that 
$l_sM^g_s \simeq \sqrt{N}-a'g^2N$.  This agrees
with the estimation made in \cite{3'}.
The coefficient $a$ in (\ref{9})
must be positive to ensure positiveness of the entropy for large 
$M_s$ \footnote{
Other reason for this is that the derivative ${dS_s\over dg} \geq 0$
that means validity of the second law in the process of collapse of the
string states.}. 

The matching  of (\ref{9}) and (\ref{6}) 
can be done only if $0\leq a <1$. This means that for large $M$ the black 
hole entropy $S_{bh}$ must grow faster than $S^g_s$ that is sensible
in the spirit of the second law. The matching condition gives the following 
critical value for the string coupling
\begin{equation}
g^2_{cr}={1\over (1-a)}{1\over (l_sM_s)} \sqrt{c\over 24}~~.
\label{10}
\end{equation} 
Equation (\ref{10}) indicates that at the matching point the correction term
in (\ref{9}) becomes important and can not be neglected. This is especially
true if the coefficient $a$ in (\ref{9}) is 
slightly different than $1$
\cite{?}.
The critical Schwarzschild radius is $r^{cr}_+=
2g^2_{cr}l^2_sM={1\over 1-a}\sqrt{c\over 6}l_s$, 
and for $a\rightarrow 1$ we have that $g^{cr},~r^{cr}_+\rightarrow \infty$.
This means that the case $a=1$ is special. Indeed, the entropies (\ref{9})
and (\ref{6}) can not be matched for any finite $g$ if $a=1$. 
However, in
this case  the ($\sim g^2$) term in Eq.(\ref{9}) exactly reproduces
the corresponding term in the black hole entropy (\ref{6}) which can be identified by  equation
\begin{equation}
S_{bh}=S^g_s-S^{g=0}_s
\label{!!}
\end{equation}
valid in the point of the transition.
Note that in arbitrary dimension $d\geq 4$ the black hole entropy 
behaves as $S^{(d)}_{bh}\simeq g^{2\over d-3} (l_sM)^{d-2\over d-3}$.
Therefore, only in four dimensions it grows as an integer power
of the string coupling $g$ that makes the interpretation (\ref{!!})
possible.

Thus, in this scenario ($a=1$) the black hole entropy appears 
as purely  {\it string-loop}
effect due to the self-interaction of the fundamental string.
Therefore, it is natural to identify the ``internal states
of black hole'' not with states of the {\it free string}
but with (a part of) states of the {\it interacting string}, and the black hole
entropy arises entirely due to the  interaction.
The total
number of the states of the interacting string
is bigger (\ref{!!}) than it is required by the black hole entropy. 
 Therefore, not all the string 
states may collapse and
form a black hole (otherwise, we would arrive at a situation
when the second law is violated). Some of them (with the entropy $S_s^{g=0}$)
must stay outside the
horizon and be accessible for an external ``observer''. 
They form gas of massless excitations (waves) 
freely propagating in a black hole. 

It is worth noting that an analogous
situation  
happens when we
are trying to calculate the quantum field theoretical entropy of
a black hole. 
The quantum entropy arises as a sum of a contribution due to the
black hole and the entropy of
the  hot gas of quantum fields propagating outside the horizon.
 The contribution of the gas can be isolated by its  dependence on 
the size $L$ of the system (see, for example, \cite{8}). 
To extract the entropy of the hole itself one 
should compare the whole quantum entropy with the entropy 
of flat space filled by the hot gas.
Possibly, the similar line of reasoning can be useful in the case
under consideration. 
Then  the contribution of flat space is due to the non-interacting
string states with the entropy $S^{g=0}_s$.

Our analysis, in principle, does not prohibit the 
appearance of the higher order terms in the expansion (\ref{9}). 
Moreover, these terms  are very likely to appear
in the regime when $a\sim 1$ and $g_{cr}$ becomes large. Formally adding \cite{?} term $b(gl_s M)^4$ to the string entropy we find that the matching condition leads to some constraint on  mass of the string forming black hole.
This happens because for sufficiently 
large mass the interacting string entropy (with higher 
order corrections added) becomes greater than the black hole entropy and the two quantities can not be matched.

In an alternative scenario
 the string interaction
appears in  such a way that the higher order corrections to (\ref{9}) do
not arise and the coefficient $a$ never becomes equal to $1$
and  changes in the limits $0\leq a<1$ only.
All next $g$-dependent corrections to the string entropy result in a 
modification of the coefficient $a$ in (\ref{9}) so that it becomes a
function of the string coupling $g$
and runs to $1$ when
$g$ goes to infinity (Eq.(\ref{10}) is an indication of this behavior)
Then, all the string states  can form   a black hole,
and the matching of the string and black hole 
entropies always can be done in accord to  the 
correspondence principle.  
At the moment, we do not have  arguments in favor of any of these
scenarios, $a> 1$, $a=1$ or $0\leq a<1$, 
and hope that further investigation will shed light on this 
problem.

The above analysis of the {\it string-black hole} correspondence
does not say us where and in which form the string states counted by
the black hole entropy present in the 
phase of the macroscopic heavy black hole. This is, however, the most
important question regarding the statistical explanation of the black 
hole entropy.
It is intuitively clear that  thin ($\sim gl_s$) layer around the horizon
is likely to be a place for storing such states \cite{1}. That is why the 
Bekenstein-Hawking entropy relates their number to geometry of the horizon.
From the analysis presented above follows that the 
{\it interaction} of the fundamental string is important for the correct
description of the {\it string-black hole} transition. Therefore,
we speculate that in the {\it black hole phase} the states in the layer 
near horizon are some {\it
bound states} due to the strong interaction of the string modes with the
horizon. These states 
arise from the modes of 
the perturbatively interacting
string in  the {\it string-black hole} transition. This picture
resembles that of given in \cite{1}. 

\section*{Acknowledgments}
I would like to thank Nemanja Kaloper for interesting discussions
and Gary Horowitz for helpful comments.
This work was  supported by NATO and NSERC.



\begin{thebibliography}
\bibitem{arg}  Since $g$ is dimensionless there may be corrections 
to the string entropy
given by
the series $\alpha_0+\alpha_1 g +\alpha_2 g^2+...$ with coefficients $\alpha_k$
 not dependent  on  parameters (mass $M$) of the system. Such 
corrections result in an additive constant to the entropy and are not of our 
interest. The corrections which are essential for us arise when $g$ comes in the combination with $l_s$, $gl_s$. This perturbation series goes with respect to the dimensionless parameter $(gl_s M)$ as shown in (\ref{9}).
\bibitem{?}  Strictly speaking, in this point the approximation (\ref{9})
is not sufficient
 and there may appear higher order corrections. Formally adding 
the term $b(gl_sM)^4$ to the expression (\ref{9}) we find that the critical
value of the string coupling becomes $g_{cr}^2={1\over 2b (l_s M)^2}(
(1-a)-\sqrt{(1-a)^2-8\pi b \sqrt{c/6}(l_sM)})$. This gives constraint on
the mass $(l_sM)\leq \sqrt{6\over c}{(1-a)^2\over 8\pi b}$. We see  that
the case $a=1$ is special even if the higher order corrections are included.
\end{thebibliography}
\end{document}